\documentstyle[12pt]{article}

\begin{document}
\pagestyle{plane}
\newcommand{\lsim}{\raisebox{0.3mm}{\em $\, <$} \hspace{-3.3mm}
\raisebox{-1.8mm}{\em $\sim \,$}}
\newcommand{\gsim}{\raisebox{0.3mm}{\em $\, >$} \hspace{-3.3mm}
\raisebox{-1.8mm}{\em $\sim \,$}}
\renewcommand{\thefootnote}{\fnsymbol{footnote}}
%%%%%%%%%%%%%%%%%%%%%%%%%%%%%%%%%%%%%%
\begin{flushright}
TIT/HEP-269/COSMO-48 \\
September \ 1994 \hspace{1.5cm} \ \\
\end{flushright}

\begin{Large}
\begin{center}
\bf{IMPLICATIONS ON THE IONIZING BACKGROUND RADIATION FROM
HST HeII GUNN-PETERSON TEST}
\end{center}
\end{Large}

\begin{center}
\begin{large}
Shin {\sc Sasaki}
\end{large}

{\it Department of Physics, Tokyo Institute of Technology,
O-okayama, Tokyo 152, Japan}

E-mail \hspace{0.5cm} sasaki@th.phys.titech.ac.jp

and

\begin{large}
Fumio {\sc Takahara}
\end{large}

{\it Department of Physics, Tokyo Metropolitan University, 
Hachioji, Tokyo 192-03, Japan}

\end{center}

\begin{center}
{\bf ABSTRACT}
\end{center}

Recently, Jakobsen et al. reported that the HeII Gunn-Peterson optical depth
is greater than 1.7 at $ z=3.3$.
When we consider this observation combined with the HI Gunn-Peterson
optical depths and the intensity
of the UV background at high redshifts, we can put constraints on the spectrum
of the UV background and the mass fraction of intergalactic medium.
We find that the spectrum must be steep
($\alpha \gsim 3; J_{\nu} \propto \nu^{-\alpha} $)
to be consistent with observations,
and that the mass fraction of the intergalactic medium must be
greater than 0.3 (0.1) in the case $ \alpha=3 (4)$ for baryon density
consistent with the primordial nucleosynthesis.
We also find that HeII fraction is as low as $10^{-3} \sim 10^{-1} $ even
for this steep spectrum.

{\it Subject headings}: Cosmology -- Intergalactic medium --
UV background:spectra
\newpage

\vspace{0.5cm}
\begin{center}
{\bf 1. Introduction}
\end{center}
\vspace{0.5cm}

Properties of intergalactic medium (IGM) at high redshifts
are an important ingredient to understand the thermal history of the universe,
and they can be probed only through absorption features in the spectra
of high redshift quasars.
The Gunn-Peterson test (GP test) which uses the Ly-$\alpha$ absorption of
atomic hydrogen, has indicated that hydrogen in the IGM
has been ionized at least since redshift $ z = 5$.
This means that the IGM is reionized after
the standard recombination epoch $ z \sim 1000$.
There are two mechanisms to ionize hydrogen: collisional ionization
and photo-ionization.
Because Cosmic Microwave Background (CMB) spectrum is very accurately given
by a black body spectrum (Mather et al. 1994),
the temperature of IGM is not high and
photo-ionization is a promising mechanism to reionize the IGM.
Then, to reionize the IGM by a photo-ionization mechanism,
one needs strong UV background radiation at high redshifts.
The existence of strong UV background radiation is also supported by
the proximity effect of Ly $\alpha$ cloud distribution (Bajtlik et al. 1988),
but the UV spectrum is not strongly constrained up to now.
This is due to the fact that we have obtained only the HI GP optical depths
and hence the UV intensity at the hydrogen Lyman edge.
However, when we get the HeII GP optical depth, we can obtain
information about its spectral shape as well as its intensity from
the HI and HeII GP optical depths.

Recently, Jakobsen et al. (1994) reported that
the HeII GP optical depth is greater than 1.7 at $ z=3.3$;
This value is larger than that was predicted in previous work
(e.g. Sasaki and Takahara 1994).
Thus, using observations of the HeII GP optical depth and of HI GP optical
depth, we can put constraints on the spectrum of UV background radiation.
Furthermore, since the ionization fraction also depends on the number
density of IGM, we expect that we can limit the IGM mass fraction.

In this Letter, we calculate the IGM thermal history with UV background
radiation, and estimate the GP optical depths of HI, HeI and HeII.
From these results, we put constraints on the spectrum of UV background
radiation and the IGM mass fraction.

\vspace{0.5cm}
\begin{center}
{\bf 2. Model and observational constraints}
\end{center}
\vspace{0.5cm}

In this Letter, for simplicity, we assume that the flux of
the background radiation is represented by

\begin{equation}
J_{\nu} (z) = \tilde{J}_{-21} \times 10^{-21} (\frac{\nu}{\nu_T})^{-\alpha}
(\frac{1+z}{4})^{\beta} {\rm ergs ~ cm^{-2} sec^{-2} Hz^{-1} sr^{-1}}
{\mbox{\hspace{0.5cm}}} (z \leq z_i),
\end{equation}
where $ \nu_T$ is the frequency at the hydrogen Lyman edge.
Here, we study the case $ \alpha=1 \sim 4 $ and $ \beta = 0 \sim 4$.
Using the proximity effect, the intensity of background radiation
is estimated as $ J_{\nu_T} = 10^{-21 \pm 0.5}
{\rm ergs ~ cm^{-2} sec^{-2} Hz^{-1} sr^{-1}} $ at $ z=1.7 \sim 3.8 $
(Bajtlik et al. 1988).
Thus, we consider a range of $ \tilde{J}_{-21} = 1/3 \sim 3$.
With this background radiation, we calculate the thermal properties
of IGM (ionization fractions of hydrogen and helium and gas temperature;
for detailed formulation, see Sasaki and Takahara, 1994).
We study the $ \Omega_0=1 $ flat universe, where $ \Omega_0 $ is the
density parameter, with the Hubble constant
$ H_0 = 50 {\rm km~sec^{-1} Mpc^{-1}} $.
We adopt the baryon density parameter $ \Omega_b = 0.05 $ and helium mass
fraction $ Y=0.25$ ($n_H : n_{He} \approx 0.08$)
to be consistent with the standard primordial nucleosynthesis
(Walker et al. 1991).
We consider only the case $ z_i=10$.
Furthermore, we assume that the mass fraction of IGM, $ f_{IGM} $,
is constant between $z=z_i$ and the present epoch.

As the observational constraints, we adopt the following observations.
(1) The HI GP optical depths
$ \tau_{GP,HI} \leq 0.18, 0.12 $ and $ 0.35 $
at $ z=2.6, 3.0 $ and $4.2$, respectively (Jenkins and Ostriker 1991).
(2) The HeI GP optical depth $ \tau_{GP,HeI} \leq 0.21 $ at $ z=1.7$
(Tripp et al. 1990),
and (3) the HeII GP optical depth $ \tau_{GP,HeII} \geq 1.7 $ at $ z=3.3$
(Jakobsen et al. 1994).
It is to be noted that
the HI and HeI GP optical depths are upper limits,
while the HeII GP optical depth is a lower limit. 
Thus, we expect that these observations should limit
the parameters to some range.
Furthermore, we consider the soft X-ray background radiation
at the present epoch
$ J_{\nu} \sim 1.3 \times 10^{-25}
{\rm ergs ~ cm^{-2} sec^{-2} Hz^{-1} sr^{-1}}$ at 1 keV (Hasinger 1992)
as an other constraint.
We regard this observed value as an upper limit
since the observed value may contain contributions from other sources
which shine only in the X-ray band.
If the assumed spectrum is flat, the soft X-ray background
becomes a strong constraint.

\vspace{0.5cm}
\begin{center}
{\bf 3. Results}
\end{center}
\vspace{0.5cm}

In Fig.1, we show the time evolution of the HI and HeII GP optical depths
for several choices of parameters.
When we consider only the HI GP optical depths, these cases are not
discriminated to each other.
For example, in Fig.1, we cannot distinguish between the case
$ (\alpha, \beta, \tilde{J}_{-21})= (1,0,1)$ and $(3,0,2)$,
because these two cases have almost the same ratio of the
HI photo-ionization coefficient to the recombination coefficient,
which results in almost the same ionization fraction of hydrogen.
On the other hand, these two cases predict different values
for the HeII GP optical depth
since  HeII photo-ionization coefficient becomes different by a factor of
$ \sim 16 $ due to a difference of spectral shape.
Thus, when we take account of both the HI and HeII GP optical depths,
we can distinguish cases which cannot be distinguished based 
only on the HI GP optical depths.

The time evolution of the GP optical depths depends on the evolution
of the UV background radiation, that is on $ \beta$.
However, since we treat a relatively small redshift range ($ z = 2.6 \sim 4.2 $
for the HI and HeII GP optical depths), the difference due to $ \beta$
is not so critical compared to other parameters.
Thus, we neglect the difference of $ \beta $ in the following discussion.

Then, we find that $ \alpha $ must be three or more in order that
predictions are consistent with all observations mentioned above.
This was pointed out by Jacobsen et al.(1994), too.
If $ \alpha$ is less than 3, strong UV background radiation which ionizes the
hydrogen in the IGM, while satisfying the observations of
the HI GP optical depths,
ionizes HeII too much, and the HeII GP optical depth becomes too small
to be consistent with the observation.
This has an important implication for the sources of the UV background.

The ionization fraction depends on the mass fraction of IGM,
$ f_{IGM} $, as well as the intensity of the
UV background radiation $ \tilde{J}_{-21} $.
For fixed value of the intensity $ \tilde{J}_{-21} $,
when $ f_{IGM} $ increases, both the number density of hydrogen and the
neutral fraction of hydrogen increase,
so the HI GP optical depth also increases.
At some critical value, it becomes to be inconsistent with the observed
upper limit.
On the other hand, when $ f_{IGM} $ decreases, both the number density
of helium and the HeII fraction decrease.
Thus, the HeII GP optical depth also decreases, and at some critical value
it becomes too small to be consistent with the observed lower limit of
the HeII GP optical depth.
Thus, when we fix the intensity $ \tilde{J}_{-21} $, we obtain an allowed
region of $ f_{IGM} $ from the HI and He II GP optical depths.
For example, in the case $ \alpha=3$ and $ \tilde{J}_{-21} =1$,
the allowed range of $ f_{IGM} $ turns out to be $ \sim 0.5 - 0.8 $.
Considering the lower limit on the intensity obtained by the proximity effect
($ \tilde{J}_{-21} \geq 1/3 $), we get $ f_{IGM} \gsim 0.3 $.
If we adopt $ \alpha = 4 $ instead of $ \alpha = 3 $ for fixed
$ \tilde{J}_{-21} $ and $ f_{IGM} $, ionization efficiency of HeI and HeII
decreases and an allowed region shifts to a region of a larger
$ \tilde{J}_{-21} $ and a smaller $ f_{IGM} $
compared with the case $ \alpha=3$.
In particular, it is more difficult to ionize HeII than to do HI and HeI,
thus the limit of the HeII GP optical depth is relaxed.
In the case $ \alpha = 4 $, we get the limit $ f_{IGM} \gsim 0.1 $
under the condition $ \tilde{J}_{-21} \geq 1/3 $.
In contrast, if $ \tilde{J}_{-21} = 3$ (Bechtold, 1994),
$ f_{IGM} \sim 1.0 (0.6 - 1.0) $ in the case $ \alpha=3 (4) $.
This conclusion is in contrast to the suggestion by Jacobsen et al. that
$ \Omega_{IGM} $ is likely to be less than $10^{-3}$.

It is to be noted that almost helium is HeIII and the fraction of
HeII is $ \sim 10^{-3} - 10^{-1} $
in the allowed region, even for a UV spectrum as steep as
$ \alpha =3 \sim 4$.
This is in contrast to the suggestion by Jacobsen et al. that HeII
may be the dominant population.

We find that all cases which satisfy the observations of the HI and
HeII GP optical depths,
are consistent with the observations of the soft X-ray radiation and the HeI GP
optical depth.
We summarize the results of the present analysis in Fig. 2.

\vspace{0.5cm}
\begin{center}
{\bf 4. Discussion and Conclusions}
\end{center}
\vspace{0.5cm}

First, we check the $z_i$ dependence.
When $z_i$ decreases, the fraction of HI and HeII decreases due to
a long recombination time, except for a short epoch near $z_i$ when
HeII is the dominant ionization state.
For $ z \lsim 4.2 $, the difference due to $z_i$ is
within a factor of a few if $ z_i \gsim 5$.
Although an allowed region in Fig.2 shifts to a smaller $\tilde{J}_{-21} $ and
a larger $ f_{IGM} $, both the limit on $ \alpha $ and the lower limit
of the fraction of IGM may not change very much.

In the above discussions, we have considered only
the $ \Omega_0 =1 $ flat universe.
When we adopt the baryon density parameter to be consistent with the
standard primordial nucleosynthesis,
the difference due to cosmological model enters only
the definition of the GP optical depth.
Consider the flat universe with cosmological constant with $ \Omega_0=0.1 $
and $ H_0 = 100 {\rm~km~sec^{-1} Mpc^{-1}} $ instead of the $ \Omega_0=1$
flat universe with $ H_0 = 50 {\rm ~km~sec^{-1} Mpc^{-1}} $.
To obtain the same GP optical depth, we need a smaller ionization fraction.
Thus, an allowed region in Fig.2 shifts towards a larger
$ \tilde{J}_{-21} $ and a smaller $ f_{IGM} $.
However, a difference due to a cosmological model is small, within
a factor of 2, and it does not change the above results qualitatively.

We find that the UV background radiation must be steep; $ \alpha \gsim 3 $
to be consistent with the observations.
If we consider absorption due to Ly $\alpha$ clouds and others,
the spectrum of background radiation will become flatter than the source
spectrum, which means that
sources must have a steeper spectrum with $ \alpha \gsim 3$.
However, the UV spectrum of quasars which are
a popular candidate of the source of UV background radiation,
is $ \sim \nu^{-1} $.
Thus, quasars are not a dominant component of source of the
UV background radiation.
We need other UV sources which have a steep spectrum.

Young galaxies are also a candidate of the source of the UV background
radiation.
If we approximate the spectrum comes from young galaxies
by a single temperature black body spectrum,
for the $\Omega_0=1$ flat universe with
$ H_0 = 50 {\rm ~km~sec^{-1} Mpc^{-1} }, f_{IGM} = 1.0$, and
$ \tilde{J}_{-21} = 1 $, we find that the temperature of black body must be
$ \sim 6 \times 10^4$ K.
This temperature is a little higher than the effective surface temperature
of O5 star.
However, this model has some difficulty in explaining the existence of highly
ionized metal absorption lines (Steidel and Sargent, 1989).

One caveat is that the observed HeII absorption trough is due to superposed
HeII line absorption in Ly $\alpha$ clouds,
which was examined by Jacobsen et al.
Although they concluded it is unlikely, this possibility is worth
a further study since the ionization state in and column density
distribution of Ly $\alpha$ clouds is not known so accurately.

In summary, we study the thermal history of IGM with a power law
UV background radiation.
We find that the spectrum of the UV background radiation must be steep
($\alpha \gsim 3 $) to be consistent with the observations.
Using the UV intensity estimated by the proximity effect,
$ f_{IGM} $ is limited as $ f_{IGM} \gsim 0.3 (0.1) $ for
$\alpha=3(4) $ for baryon density
consistent with the primordial nucleosynthesis.
We also find that HeII fraction is as low as $10^{-3} \sim 10^{-1} $ even
for this steep spectrum.

\vspace{0.5cm}

%{\bf Acknowledgment}

\vspace{0.5cm}
\begin{center}
{\bf References}
\end{center}
\vspace{0.5cm}

\def\pp{\par\parshape 2 0truecm 16.5truecm 1truecm 15.5truecm\noindent}
\def\apjpap#1;#2;#3;#4;#5;#6; {\pp#1, #2, {#3} {{#4}}, {#5}. {#6}}
\def\apjbook#1;#2;#3;#4;#5; {\pp{#1}, #2, {\it #3} (#4: #5)}
\def\apjppt#1;#2;#3; {\pp#1, #2, #3.}
\def\apjproc#1;#2;#3;#4;#5;#6; {\pp#1, #2, in {\it #3}, #4, (#5), #6.}

\apjpap Bajtlik, S., Duncan, R. and Ostriker, J.P.;1988;ApJ;327;570;;
\apjpap Bechtold, J.;1994;ApJS;91;1;;
\apjproc Hasinger, G.;1991; X-ray Emission from AGN and the Cosmic X-ray
      Background;eds W.Brinkmann \& J. Tr\"{u}mper;Garching:MPE;321;
\apjpap Jakobsen, P. et al.;1994;Nature;370;35;;
\apjpap Jenkins, E.B. and Ostriker, J.P.;1991;ApJ;376;33;;
\apjpap Mather, J.C. et al.;1994;ApJ;420;439;;
\apjpap Sasaki, S. and Takahara, F.;1994;Prog.Theor.Phys.;91;699;;
\apjpap Steidel, C.C. and Sargent, W.L.W.;1989;ApJ;343;L33;;
\apjpap Tripp, T.M. et al.;1990;ApJ;364;L29;;
\apjpap Walker, T.P. et al.;1991;ApJ;376;51;;

\vspace{1cm}

\begin{center}
{\bf Figure Captions}
\end{center}

\noindent
Figure 1 {\hspace{0.5cm}}
GP optical depths plotted against $ z$ for (a) HI and (b) HeII.
The solid, dotted and dashed lines represent
$ (\alpha, \tilde{J}_{-21}) = (1,1), (3,1) $ and $(3,2)$, respectively.
The arrows represent the limit due to GP test.
Here, we fix $ f_{IGM} = 1.0 $.

\noindent
Figure 2 {\hspace{0.5cm}}
Allowed region in the $(\tilde{J}_{-21}, f_{IGM} )$ plane.
The upper boundary comes from the HI GP test and the lower boundary comes from
the HeII GP test.

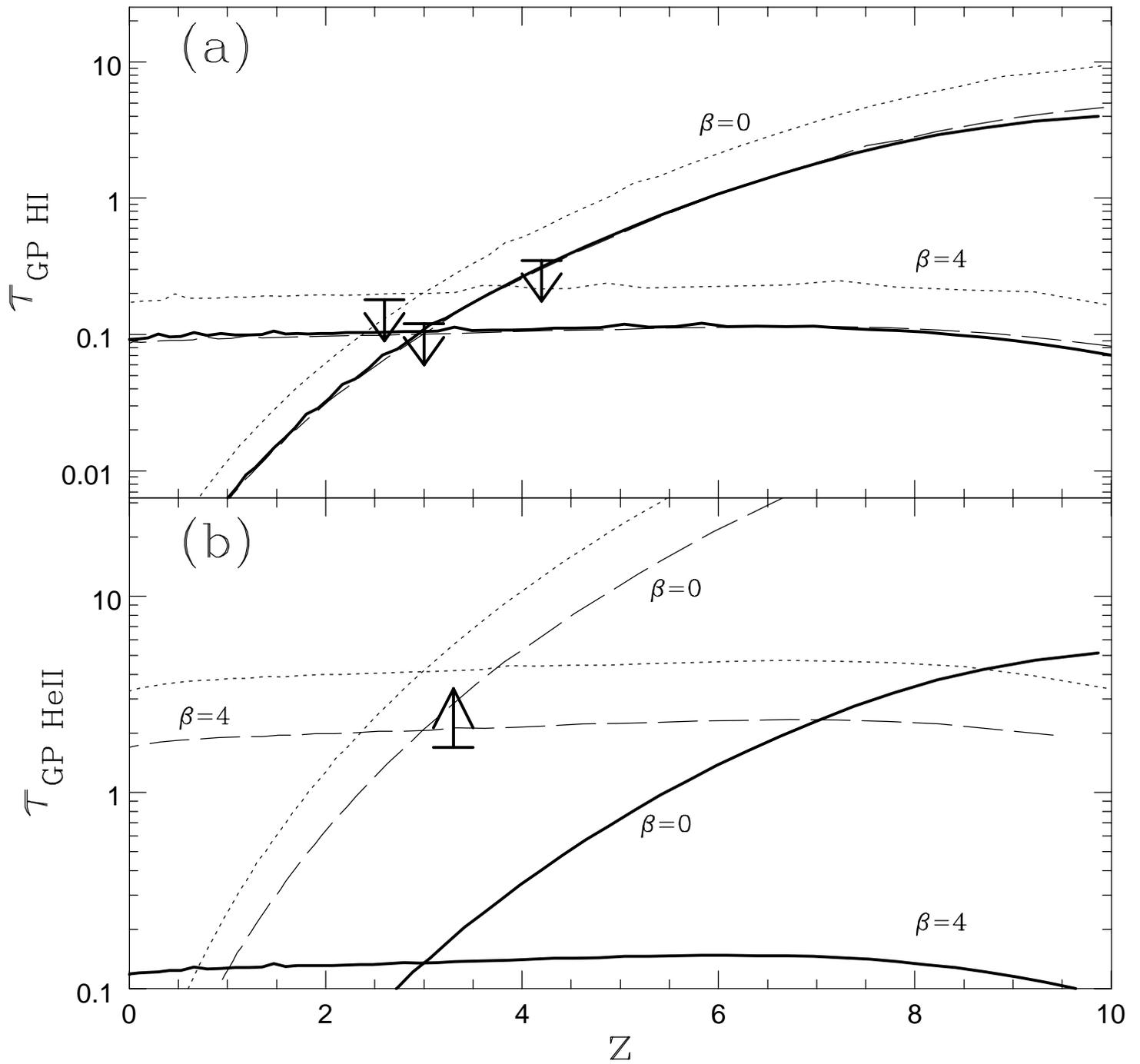
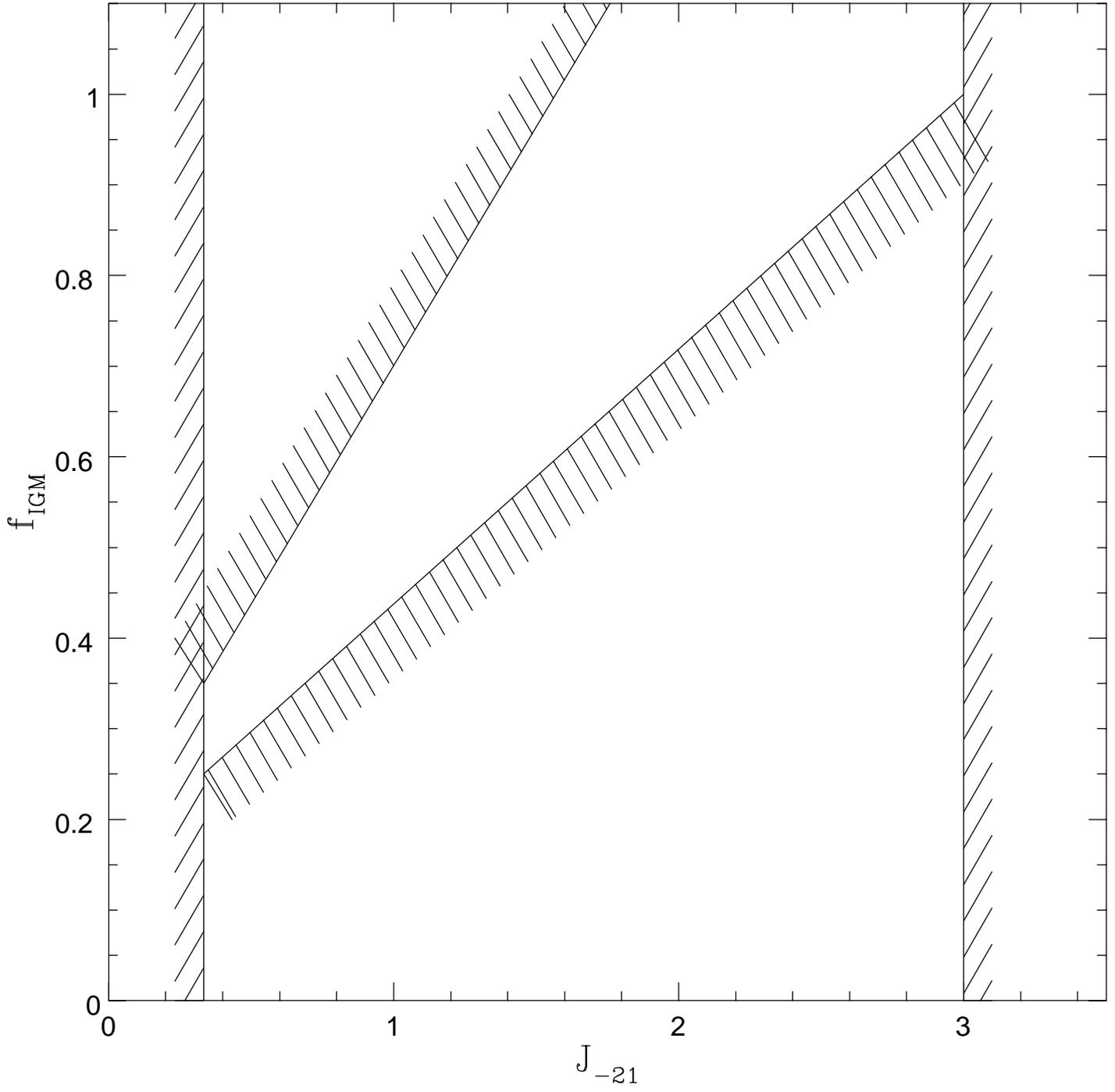
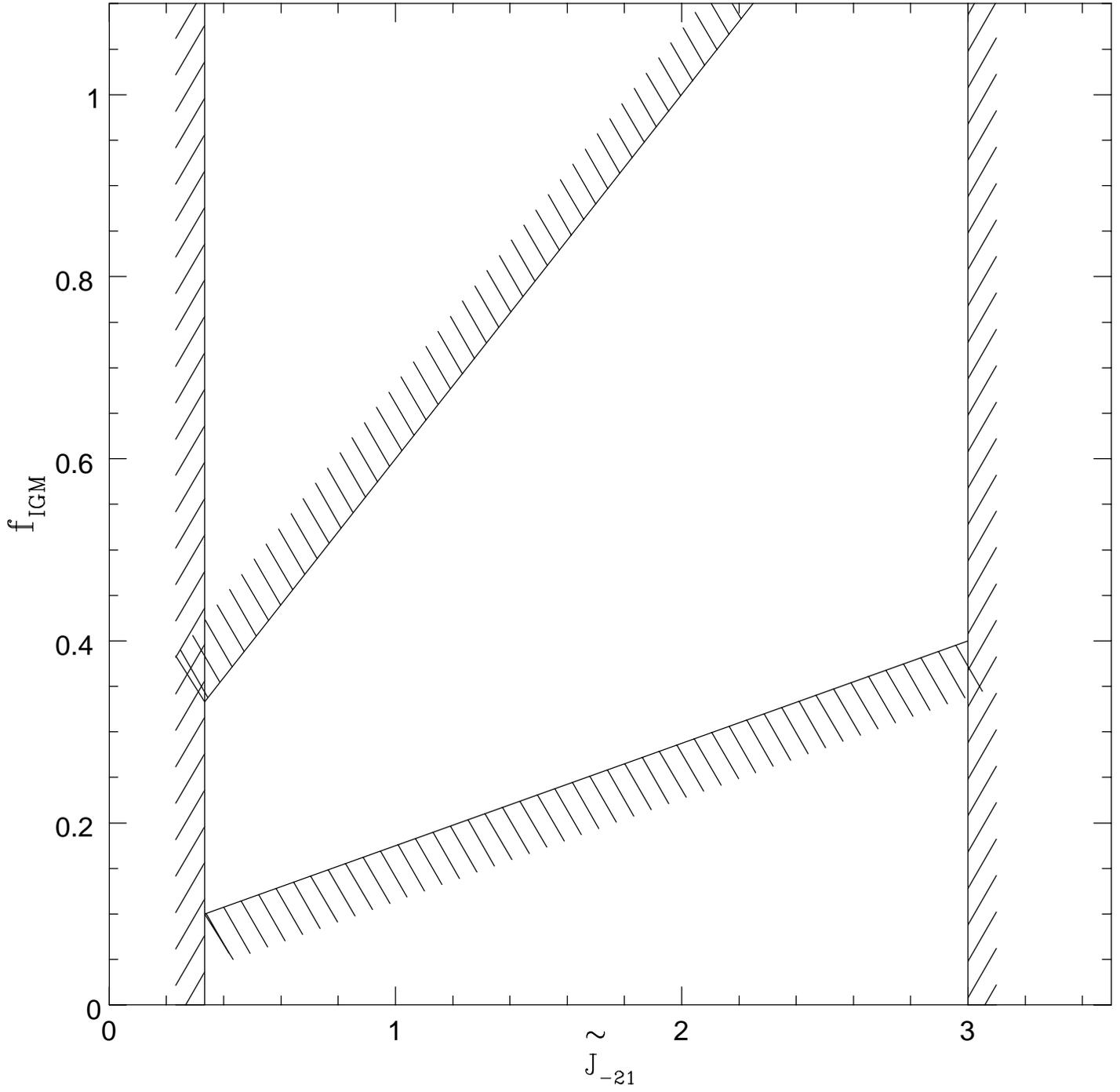
\end{document}